\documentclass[onecollarge]{svjour2}
\usepackage{bm}
\usepackage{amsmath}
\usepackage{amssymb}
\usepackage{graphicx}           
\usepackage{epsfig}
\usepackage{color}

\newcommand\nn{\nonumber}
\newcommand\bea{\begin{eqnarray}}
\newcommand\eea{\end{eqnarray}}
\newcommand\f{\frac}
\newcommand\p{\partial}
\newcommand\la{\langle}
\newcommand\ra{\rangle}

\begin{document}

\title{Tagged particle diffusion in one-dimensional gas with Hamiltonian dynamics}

\author{Anjan Roy\and Onuttom Narayan \and Abhishek Dhar \and Sanjib Sabhapandit}

\institute{A. Roy \at Raman Research Institute, Bangalore 560080, India
\and O. Narayan \at Department of Physics, University of California, Santa Cruz, California 95064, USA
\and A. Dhar \at International Centre for Theoretical Sciences, TIFR, Bangalore 560012, India
\and S. Sabhapandit \at Raman Research Institute, Bangalore 560080, India
}
\date{\today}

\maketitle

\begin{abstract}
We consider  a one-dimensional gas of hard point particles in a finite box  that are in thermal equilibrium and  evolving under Hamiltonian dynamics. Tagged particle correlation functions of the middle particle are studied. For the special 
case where all particles have the same mass, we obtain analytic results for the 
velocity auto-correlation function  in the short time diffusive regime and the long time approach to the saturation value when finite-size effects become relevant. In the case where the 
masses are unequal,  numerical simulations indicate sub-diffusive behaviour with mean square displacement of the tagged particle growing  as $ t/\ln (t)$ with time $t$. 
Also  various correlation functions, involving the velocity and position of the tagged particle, show damped oscillations at long times that are absent for the equal mass case.
\end{abstract}

\section{Introduction}

Observing the dynamics of a single tagged particle in a many particle
system constitute a simple way of probing the complex dynamics of an
interacting many body system and has been studied both theoretically~\cite{jepsen65,harris65,lebowitz67,lebowitz72}
and experimentally~\cite{hahn96,wei00,lutz04}. Much of the theoretical studies on tagged particle
diffusion have focussed on one-dimensional systems and discussed two
situations where the microscopic particle dynamics is (i) Hamiltonian~\cite{jepsen65,lebowitz67,lebowitz72}
or (ii) stochastic~\cite{harris65,beijeren83,kollman03,lizana08,gupta07,barkai09,barkai10}.

For systems with Hamiltonian dynamics, the evolution of the system is
completely deterministic and all the randomness in the system is due
to the randomness in the initial condition.  One of the earliest
result for systems with Hamiltonian evolution is that of
Jepsen~\cite{jepsen65} on tagged particle diffusion in a
one-dimensional hard particle gas of elastically colliding particles
of equal masses.  For an infinite system at a fixed density of
particles Jepsen showed that the mean square deviation (MSD) of a
tagged particle from its initial position grows linearly with time
$t$. He obtained an explicit expression for the diffusion constant and
the related velocity autocorrelation function (VAF).  This was done by
exploiting the fact that when two particles of equal mass collide
elastically in a one-dimensional system, their velocities are
exchanged; if we ignore tags on particles, this is is equivalent to
the particles passing through each other without colliding,
simplifying the dynamics.

For a finite system, with $N$ particles, there must be corrections
to Jepsen's result, since the MSD must saturate at long time (to a
value that depends on the size of the system). This situation has been 
extensively studied for stochastic dynamics \cite{beijeren83,kollman03,lizana08,gupta07,barkai09,barkai10} but not much for the Hamiltonian 
case~\cite{evans79,kasper85,lebowitz72}. Lebowitz and Sykes~\cite{lebowitz72} considered finite size effects for 
some special initial conditions. In this paper we consider Boltzmann 
distributed initial coniditions.  The first objective
of this paper is to obtain analytical expressions for the 
VAF that are valid over the entire regime: both $t << N$ (\`a la
Jepsen) and $t >> N.$ 

If the particle masses are not all the same in a hard particle gas,
there are no analytical results. Since the dynamics are expected
to be ergodic, the correlation functions should be very different
from those of the equal mass particle gas. Indeed, a simulation
study~\cite{marro85} of a gas where odd and even numbered particles have different
masses suggested that the decay of the VAF with time $t$ in this
model was as $ \sim t^{-\delta}$ with $\delta \lesssim 1$ which is
completely different from the Jepsen result ($\sim t^{-3}$). If
this is correct, it would imply that tagged particle motion is
superdiffusive in this system.
The second objective of this paper is to accurately obtain the decay
of the VAF and other  correlation functions for a hard 
particle gas with unequal masses, to see if
tagged particle motion is superdiffusive. We perform simulations
on a one-dimensional gas with alternating masses. To ascertain how robust the numerical results are we also do simulations  with randomly chosen
masses.

Although there has been considerable work on the (hydro)dynamics
of one dimensional hard particle gas and other systems in the context
of heat conduction~\cite{narayan02}, this involves the propagation of conserved
quantities as a function of position and time without reference to
the identity of each particle. This changes things considerably:
for instance, conserved quantities propagate ballistically for an
equal mass hard particle gas, resulting in a thermal conductivity
proportional to $N,$ while tagged particle dynamics in the same
system is diffusive.  Thus here we approach the dynamics from a
perspective that is different from the heat conduction literature.

In sec.~(\ref{sec:analytic}) we define the model and dynamics and give analytic results for the VAF in the special case where all masses are equal. In sec.~(\ref{sec:simulations}) we  present the simulation results for the VAF and other correlation functions for the general case where masses are not all equal. 
 We summarize our results in Sec.~\ref{sec:summary}. Some details of the
calculation are given in Appendix~\ref{appendix}.

\section{Analytic results for equal mass hard-particle gas}
\label{sec:analytic}

Here we consider a gas of $N=2M+1$ point particles in a one-dimensional
box of length $L$. The particles interact with each other through
hard collisions conserving energy and momentum. The Hamiltonian of
the system thus consists of only kinetic energy. All the particles
have the same mass $m.$ In any interparticle collision, the two
colliding particles exchange velocities. When a terminal particle
collides with the adjacent wall, its velocity is reversed. The
initial state of the system is drawn from the canonical ensemble
at temperature $T$.  Therefore, the initial positions of the particles
are uniformly distributed in the box.  Let $x_i$ be the position
of the $i$-th particle measured with respect to the ``left'' wall,
and $0 < x_1 < x_2 < \dotsb < x_{N-1} < x_N < L$.  The initial
velocities of the particles are choosen independently from the
Gaussian distribution with zero mean and a variance $\overline{v}^2=k_B
T/m$.

By exchanging the identities of the  particles
emerging from  collisions, one can effectively treat the system as 
non-interacting~\cite{jepsen65}.
In the non-interacting picture, each particle executes an independent
motion.  The particles pass through each other when they `collide' and
reflect off the walls at $x=0$ and $x=L.$ The initial condition is
that each particle is independently chosen from the single particle
distribution $p(x,v)=L^{-1} (2 \pi
\bar{v}^2)^{-1/2}e^{-v^2/2\bar{v}^2}$, where $\bar{v}^2=k_B T/m$.  To
find the VAF of the middle particle in the interacting-system from the
dynamics of the non-interacting system, we note that there are two
possibilities in the non-interacting picture: (1) the same particle is
the middle particle at both times $t=0$ and $t$, or (2) two different
particles are at the middle position at times $t=0$ and $t$
respectively. We denote the VAF corresponding to these two cases by
$\langle v_M(0) v_M(t)\rangle_1$ and $\langle v_M(0) v_M(t)\rangle_2$
respectively. The complete VAF is given by $\la v_M(0) v_M(t) \ra=\la v_M(0) v_M(t) \ra_1+\la v_M(0) v_M(t) \ra_2$. 
We now present a physically motivated derivation of these two 
quantities. A   direct derivation and some more details are given in an 
appendix.

We first define a few quantities. The probability density
for a (non-interacting) particle to be at $x$ at time $t=0$ and $y$ at 
time $t$ is 
\begin{align}
P(x, y; t) &= \frac{1}{L\sqrt{2\pi} \overline v t}
\sum_{n=-\infty}^\infty
\Biggl\{\exp\left[-\frac{(2 n L + y - x)^2}{2\overline v^2 t^2}\right] 
+ \exp\left[-\frac{(2 n L - y - x)^2}{2\overline v^2
    t^2}\right]\Biggr\}\nonumber\\
&= \frac{1}{L^2} \sum_{k=-\infty}^\infty \cos\frac{\pi k x}{L}\cos\frac{\pi k y}{L} \exp[-\overline v^2 t^2 k^2 \pi^2/(2 L^2)].
\label{pxy}
\end{align}
The first line is easily obtained by realizing that, for a free
particle in an infinite box with a Gaussian velocity distribution,
the corresponding probability density is $(L\overline v t
\sqrt{2\pi})^{-1}\exp[- (x - y)^2/(2\overline v^2 t^2)],$ and the
boundaries at $x=0$ and $L$ set up an infinite sequence of image
sources. The second line is obtained using the Poisson resummation formula
 or by realizing that with $\tau=t^2,$ the first expression satisfies $\partial_\tau P = - \overline v^2 \partial_y^2 P$ with initial condition $P(x, y, 0) = \delta(x - y)/L$ and boundary conditions $\partial_y P(x,0, \tau) = \partial_y P(x, L, \tau) = 0,$ and expressing this in terms of the eigenfunctions of the Laplacian.
As a variant of Eq.~(\ref{pxy}) we also define the function 
\begin{align}
P_-(x, y; t) &=
\frac{1}{L\sqrt{2\pi}\overline v t}\sum_{n=-\infty}^\infty
\biggl\{\exp\left[-\frac{(2 n L + y - x)^2}{2\overline v^2 t^2}\right]
- \exp\left[-\frac{(2 n L - y - x)^2}{2\overline v^2
    t^2}\right]\biggr\}\nonumber\\
&= \frac{1}{L^2} \sum_{k=-\infty}^\infty \sin\frac{\pi k x}{L}\sin\frac{\pi k y}{L} \exp[-\overline v^2 t^2 k^2 \pi^2/(2 L^2)].
\label{P-}
\end{align}

In terms of these functions, the correlation function $\langle v_M(0)
v_M(t)\rangle_1$ can be found by picking one of the non-interacting
particles at random, calculating the probability that it goes from
$(x, 0)$ to $(y,t)$ and that it is in the middle at both $t=0$ and
$t,$ multiplying by $v(0) v(t)$ and integrating over $x$ and $y.$ The
multiplication by $v(0) v(t)$ is equivalent to inserting a factor of $(2 n
L + y - x)^2/t^2$ in the first term of the first line of Eq.(\ref{pxy})
and a factor of $-(2 n L - y - x)^2/t^2$ in the second term, since
they correspond to even and odd number of reflections respectively.
Thus one obtains the normalized correlation function (see appendix)
\begin{equation}
C_{vv}^{(1)} = \f{\la v_M(0) v_M(t) \ra_1}{\bar{v}^2} =  
N \int_{0}^{L} dx \int_{0}^{L} dy
~P_{N}^{(1)}(x, y, t)
~\partial_{\overline v} [\overline v P_-(x,y,t)],
\label{jep1-1}
\end{equation} 
where $P_N^{(1)}(x, y, t)$ is the probability that there are an equal
number of particles to the left and right of $x$ and $y$ at $t=0$ and $t$
respectively.

Turning to $\langle v_M(0) v_M(t)\rangle_2,$ we pick two particles at
random at time $t=0,$ calculate the probability that they go from $(x,0)$
to $(y,t)$ and $(\tilde x, 0)$ to $(\tilde y, t),$ that there are an
equal number of particles on both sides of $x$ and $\tilde y$ at $t=0$
and $t$ respectively, multiplying by $v(0)\tilde v(t)$ and integrating
with respect to $x,y,\tilde x, \tilde y$.  From Eq.(\ref{pxy}), multiplying
$P(x, y, t)$ by $v(0)$ is equivalent to inserting a factor of $(2 n L +
y - x)/t$ and $(2 n L - y - x)/t$ in front of the first and second terms
respectively in the first line. Also, multiplying $P(\tilde x, \tilde y,
t)$ by $\tilde v(t)$ is equivalent to inserting a factor of $(2 n L +
\tilde y - \tilde x)/t$ and $- (2 n L - \tilde y - \tilde x)/t$ in front
of the two terms. Converting these factors to appropriate derivatives,
we have for the normalized correlation function
\begin{equation}
C_{vv}^{(2)} = N(N-1)
\idotsint dx d\tilde x 
dy d\tilde y  ~P_{N}^{(2)}(x,\tilde x, y, \tilde y, t) 
\bigl[\overline v t~\partial_x P(x,y, t)\bigr] 
\bigl[-\overline v t~ \partial_{\tilde y}  P(\tilde x,\tilde
  y)\bigr], \label{jep2-1}
\end{equation}
where $P^{(2)}_N(x,y, \tilde x,\tilde y, t)$ is the probability that
there are an equal number of particles on both sides of $x$ and
$\tilde y$ at $t=0$ and $t$ respectively, given that there is one
particle at $(\tilde x, 0)$ and at $(y, t).$

To proceed further, we need the expressions for $P^{(1,2)}_N$.  For this we define $p_{-+}(x, y; t)$ as the probability that a particle
is to the left of $x$ at $t=0$ and to the right of $y$ at time $t$. Let
$p_{+-}$, $p_{--}$ and $p_{++}$ be similarly defined.  Thus
\bea p_{-+}(x,y;t)=\int_{0}^x dx' \int_y^{L} dy' P(x',y';t)~, \nn \\ 
p_{+-}(x,y;t)=\int_x^{L} dx' \int_{0}^y dy' P(x',y';t) ~, \nn \\ 
p_{--}(x,y;t)=\int_{0}^x dx' \int_{0}^y dy' P(x',y';t)~, \nn \\ 
p_{++}(x,y;t)=\int_x^{L} dx' \int_y^{L} dy' P(x',y';t)~. 
\label{transp} \eea 
In terms of the expressions defined in Eqs.(\ref{transp}),
it is straightforward to see that 
\begin{align}
P_{N}^{(1)}(x,y, t) &=
\int_{-\pi}^{\pi} \frac{d\phi}{2\pi}\int_{-\pi}^{\pi}\frac{d\theta}{2\pi} 
~\Bigl[p_{++} e^{i\phi} + p_{--} e^{-i\phi} + p_{+-} e^{i\theta} + p_{-+} e^{-i\theta}\Bigr]^{N-1} \nn \\
&=\f{2}{(2 \pi)^2} \int_{-\pi/2}^{\pi/2} d\phi \int_{-\pi}^{\pi} d\theta 
~\Bigl[1-(1-\cos{\phi}) ~(p_{++}+ p_{--}) 
+i \sin \phi ~(p_{++}-p_{--}) \nn \\
& ~~~~~~~~~~~~~~~~~~~~~~~~~~~~~~~~~~~-(1-\cos {\theta})~ 
(p_{+-} + p_{-+} ) 
+i \sin \theta~(p_{+-}-p_{-+}) \Bigr]^{N-1}~, 
\label{jep1-3}
\end{align}
where we used the identity $p_{++}+p_{+-}+p_{-+}+p_{--}=1$, the fact that $N-1$ is even and the integrands are
unchanged if $\theta,\phi$ are increased by $\pi$.
The angular integrals enforce the conditions that if $m + n$ particles
are to right of $x$ at time $t=0,$ of which $n$ particles cross from
right to left in time $t,$ then $n$ particles cross from left to right
and $m$ particles remain on the left, so that the number of particles
on both sides of $x$ at time $t=0$ and $y$ at time $t$ is $m+n$ (see appendix for more details). From 
Eqs.(\ref{jep1-1}) and (\ref{jep1-3}) we get
\begin{align}
C_{vv}^{(1)} &=
 N \f{2 }{(2\pi)^2}
\int_0^L dx \int_0^L d y \int_{-\pi/2}^{\pi/2} d\phi \int_{-\pi}^\pi d\theta 
~ \p_{\bar{v}} [\overline v P_-(x,y, t)] \nn \\
&\qquad \times \Bigl[1-(1-\cos{\phi}) ~(p_{++}+ p_{--})+i \sin \phi ~(p_{++}-p_{--})
\nn\\
&\qquad\qquad-(1-\cos {\theta})~ 
(p_{+-} + p_{-+} )  +i \sin \theta~(p_{+-}-p_{-+}) \Bigr]^{N-1}~. 
\label{v0vt1}
\end{align}

Using similar arguments as used for Eq.(\ref{jep1-3}) (see appendix), 
one can write 
\begin{align}
P_{N}^{(2)}(x,y, \tilde x, \tilde y, t) &= \int_{-\pi}^\pi 
\frac{d\phi}{2\pi} \int_{-\pi}^\pi \frac{d\theta}{2\pi}
~\Bigl[p_{++}(x,\tilde y) e^{i\phi} + p_{--}(x,\tilde y) e^{-i\phi} \nn\\
&\qquad\qquad \qquad
+ p_{+-}(x,\tilde y) e^{i\theta} + p_{-+}(x,\tilde y)
  e^{-i\theta}\Bigr]^{N-2}\, \psi(\theta,\phi) \nn \\
&=\f{2}{(2 \pi)^2} \int_{-\pi/2}^{\pi/2} d\phi \int_{-\pi}^{\pi} d\theta 
~\Bigl[1-(1-\cos{\phi}) ~(p_{++}+ p_{--})\nn\\
&\qquad\qquad 
+i \sin \phi ~(p_{++}-p_{--}) -(1-\cos {\theta})~ 
(p_{+-} + p_{-+} ) \nn\\
&\qquad\qquad \qquad
 +i \sin \theta~(p_{+-}-p_{-+}) \Bigr]^{N-2}~\psi(\theta,\phi)~.
\label{jep2-2}  
\end{align}
where
\begin{equation}
\psi(\theta,\phi)= 
\begin{cases}
e^{-i \phi} &\text{for}~~ x > \tilde x~, y < \tilde y  \\
 e^{-i \theta} &\text{for}~~ x > \tilde x~,~y  > \tilde y  \\
 e^{i \phi} &\text{for}~~ x < \tilde x~,~y > \tilde y \\
 e^{i \theta} &\text{for}~~ x < \tilde x~,~ y < \tilde y~.
\end{cases}
\end{equation}
Using the second line of Eq.(\ref{pxy}) for $P(x, y, t)$ and $P(\tilde x, \tilde y, t),$ 
integrating over $y$ and $\tilde x,$  and comparing to the second line of Eq.(\ref{P-}) we
obtain 
\begin{multline}
C_{vv}^{(2)} =
 -N (N - 1) \overline v^2 t^2\f{2 }{(2\pi)^2}
\int_0^L dx \int_0^L d\tilde y \int_{-\pi/2}^{\pi/2} d\phi \int_{-\pi}^\pi d\theta 
~ P_-^2(x,\tilde y, t)   (2\cos\phi - 2\cos\theta)\\
\times \Bigl[1-(1-\cos{\phi}) ~(p_{++}+ p_{--})+i \sin \phi ~(p_{++}-p_{--})
-(1-\cos {\theta})~ 
(p_{+-} + p_{-+} )  +i \sin \theta~(p_{+-}-p_{-+}) \Bigr]^{N-2}~.
\label{v0vt2}
\end{multline}

\subsection{Short time regime}
\begin{figure}[t]
\begin{center}
\includegraphics[width=0.7\hsize]{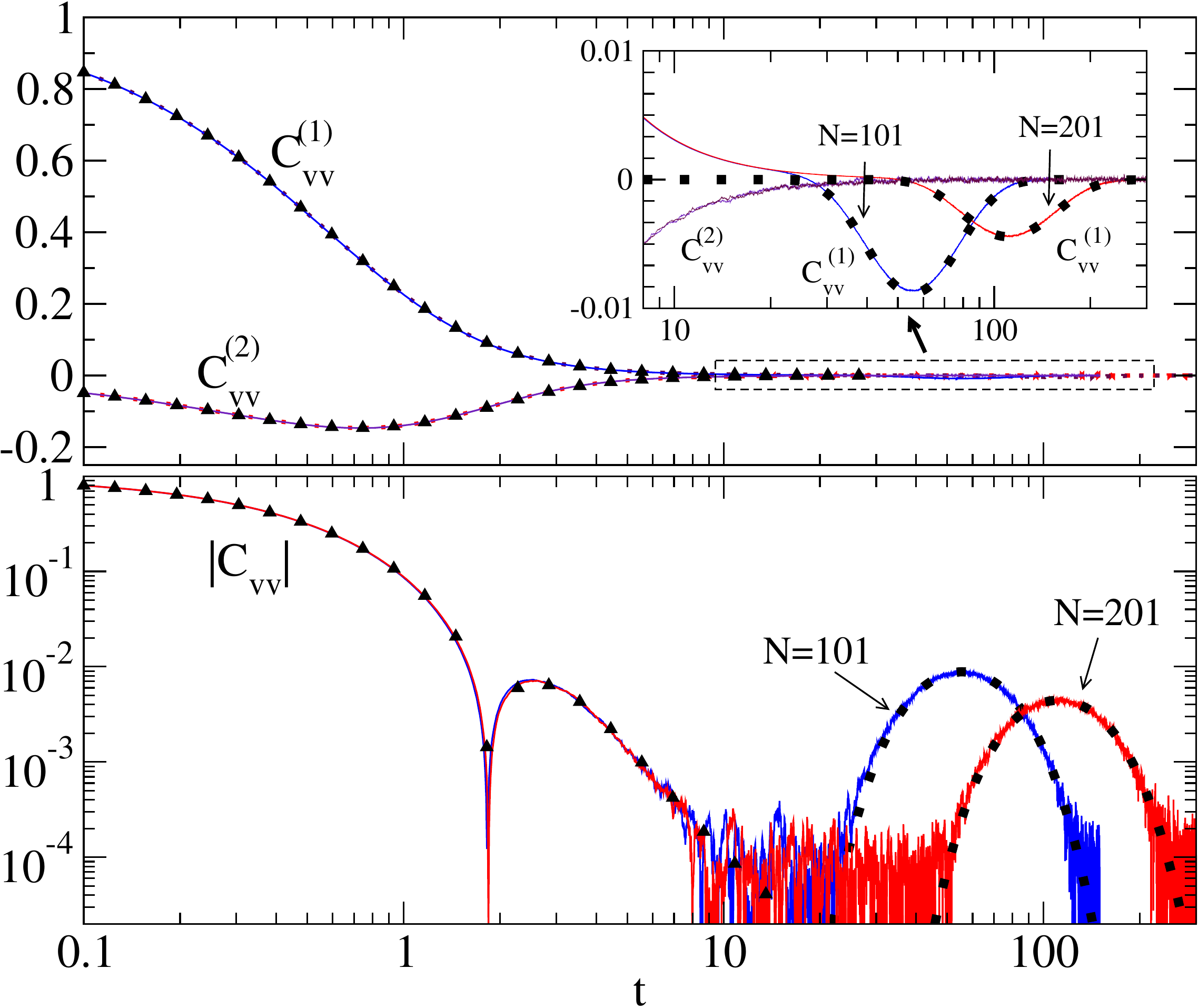}
\caption {\label{eqmC1C2}Plot of the separate contributions
  $C_{vv}^{(1)}$ and $C_{vv}^{(2)}$ to the velocity-autocorrelation
  function of equal mass hard-particle gas, for two different system
  sizes ($N=101,201$) with fixed density $\rho=1$, and $\overline v =1$. The solid lines correspond to the simulation data, whereas the
  points are from the analytical results given by
  Eqs.~\eqref{v0vt1} and \eqref{v0vt2} for short time behaviors
  ($\blacktriangle$), and Eq.~\eqref{C-long} for the long time
  behaviors ($\blacksquare$).  In the first panel, the dashed rectangle is enlarged in the inset.}
\end{center}
\end{figure}

When $\bar{v}t << L,$ the tagged particle does
not feel the effect of the walls and we can make the following
approximations \bea P(x, y, t) = P_-(x, y, t) =\frac{1}{\sqrt{2\pi}\overline v t L}\exp\left[-\f{(y-x)^2}{2 \bar{v}^2
    t^2}\right]~. \label{Sm1} \eea In this limit the expressions for
$p_{-+}$, {\emph {etc.}} given in Eq.~\eqref{transp} also simplify by
using Eq.~\eqref{Sm1} and taking the limits of the $y'$ integral to be
from $y$ to $\infty$ for $p_{-+},p_{++}$ and from $-\infty$ to $y$ for
$p_{+-},p_{--}$. We then get
\begin{align}
Np_{++}(x,y;t) &\approx \f{\rho}{\sqrt{2 \pi} \bar{v} t}\int_x^{L} dx'
\int_y^{\infty} dy' ~e^{-{(y'-x')^2}/{2 \bar{v}^2 t^2}} ~
\approx\f{N}{2}-\f{\lambda z_+}{2}-\lambda q(z_-)~,\nn  \\
Np_{--}(x,y;t) &\approx \f{\rho}{\sqrt{2 \pi} \bar{v} t} \int_{0}^x
dx' \int_{-\infty}^y dy'~ e^{-{(y'-x')^2}/{2 \bar{v}^2
    t^2}}\approx\f{N}{2}+\f{\lambda z_+}{2}-\lambda q(z_-)~, \nn \\
Np_{+-}(x,y;t) &\approx \f{\rho}{\sqrt{2 \pi} \bar{v} t} \int_x^{L}
dx' \int_{-\infty}^y dy' ~e^{-{(y'-x')^2}/{2 \bar{v}^2 t^2}}
\approx-\f{\lambda z_-}{2}+\lambda q(z_-)~, \nn \\
Np_{-+}(x,y;t) &\approx \f{\rho}{\sqrt{2 \pi} \bar{v} t} \int_{0}^x dx'
\int_y^{\infty} dy' ~e^{-{(y'-x')^2}/{2 \bar{v}^2 t^2}} \approx \f{\lambda
  z_-}{2}+\lambda q(z_-)~,  \label{transp1} \\
{\rm where}~z_+ &=\f{x + y-L}{\bar{v} t},~ z_-=\f{x-y}{\bar{v}t}~,\lambda= \rho \bar{v} t~~\text{with}~ \rho = N/L,\nn \\
{\rm and}~q(z_-)&=\f{e^{-z_-^2/2}}{\sqrt{2 \pi}}  +\f{z_-}{2}
\mbox{Erf}(z_-/\sqrt{2})~
\end{align}
Hence we get $N(p_{++}+p_{--})=N-2 \lambda
q(z_-),~N(p_{++}-p_{--})=-\lambda z_+,~N(p_{+-}+p_{-+})=2 \lambda
q(z_-),~N(p_{+-}-p_{-+})=-\lambda z_-$~.  Using these in
Eqs.~(\ref{v0vt1}) and (\ref{v0vt2}), and changing variables from
$x_0,x_t$ and $x_0,\tilde{x}_t$ to $z_+,z_-$ we get, for large $N$
\begin{align}
C_{vv}^{(1)}(t)&= \f{\lambda}{\sqrt{2 \pi} (2 \pi)^2 } \int_{-\infty}^\infty
dz_+ \int_{-\infty}^\infty dz_- z_-^2 e^{-z_-^2/2} 
\int_{-\pi/2}^{\pi/2} d \phi \int_{-\pi}^{\pi} d \theta\,
\nn \\ & \qquad\qquad
\times  
e^{-N(1-\cos \phi)} e^{-i \lambda z_+ \sin \phi} e^{-2 \lambda q(z_-) (1-\cos \theta)} e^{-i \lambda z_- \sin \theta}~, \nn \\
C_{vv}^{(2)}(t) &= -\f{\lambda^2}{4 \pi^3} \int_{-\infty}^\infty
dz_+ \int_{-\infty}^\infty dz_-  e^{-z_-^2} 
\int_{-\pi/2}^{\pi/2} d \phi \int_{-\pi}^\pi d \theta
~(\cos \phi -\cos \theta)~
\nn \\ & \qquad\qquad\times 
e^{-N(1-\cos \phi)} e^{-i \lambda z_+ \sin \phi} e^{-2 \lambda q(z_-) (1-\cos \theta)} e^{-i \lambda z_- \sin \theta}~. \nn
\end{align}
For large $N$, the major contribution of the integral over $\phi$
comes from the region around $\phi=0$. Therefore, the $\phi$ integral
can be performed by expanding around $\phi=0$ to make it a Gaussian
integral (while extending the limits to $\pm\infty$). Subsequently,
one can also perform the Gaussian integral over $z_+$. This leads to
the following expressions:
\begin{align}
\label{C1-short}
 C_{vv}^{(1)}(t) &= \f{1}{(2 \pi)^{3/2}}
 \int_{-\infty}^\infty dz_- z_-^2 e^{-z_-^2/2}
 \int_{-\pi}^\pi d \theta  e^{-2 \lambda q(z_-) (1-\cos \theta)} e^{-i
   \lambda z_- \sin \theta}~, \\
\label{C2-short}
 C_{vv}^{(2)}(t) &= -\f{\lambda}{2\pi^2}
 \int_{-\infty}^\infty dz_-  e^{-z_-^2} ~ \int_{-\pi}^\pi d \theta
~(1 -\cos \theta)~  e^{-2 \lambda q(z_-) (1-\cos \theta)} e^{-i \lambda z_- \sin \theta}~.
\end{align}
Thus we have closed form expressions of the VAF which are valid in the entire 
short time regime. 
For any value of $\lambda=\rho \overline v t $, these integrals can be performed
numerically, and as we see from Fig.~\ref{eqmC1C2}, the results are in
excellent agreement with the numerical simulation.

We now analyze the above expression in the large $\lambda$ limit,
i.e., $(\overline v \rho)^{-1} \ll t.$ Together with the condition $\overline v t << L$ for being in the short time
regime, this means that $t$ is much larger than the typical time between interparticle collisions,
i.e. outside the ballistic regime, while being much smaller than the time it takes to see finite size
effects. We first make
a change of variables $\sqrt\lambda z_{-}=z$ and
$\sqrt\lambda\theta=x$. The integrands can then be expanded as a power
series in powers of $1/\lambda$. The integrals acquire the forms:
\begin{align}
 C_{vv}^{(1)}(t) &= \f{1}{(2 \pi)^{3/2}} 
 \int_{-\infty}^\infty dz\int_{-\sqrt\lambda\pi}^{\sqrt\lambda\pi} dx\,
e^{-\frac{x^2}{\sqrt{2 \pi }}-i x z}\sum_{n=2}^\infty a_n(x,z)\lambda^{-n},
\\ 
 C_{vv}^{(2)}(t) &= -\f{1}{2\pi^2} 
 \int_{-\infty}^\infty dz \int_{-\sqrt\lambda\pi}^{\sqrt\lambda\pi} dx\,
e^{-\frac{x^2}{\sqrt{2 \pi }}-i x z}
\sum_{n=1}^\infty b_n(x,z)\lambda^{-n},
\end{align}
where $a_n(x,z)$ and $b_n(x,z)$ are polynomials in $x$ and $z$. For
example, $a_2(x,z)=z^2$, $b_1(x,z)=x^2/2$, and so on. Now, integrating
term by term (while extending the integrating limits of $x$ to
$\pm\infty$) we get
\begin{align}
C_{vv}^{(1)}(t)&=\frac{1}{\pi}\lambda^{-2}
-\frac{2(\pi-3 )}{(2\pi)^{3/2}} \lambda^{-3} +O(\lambda^{-5}),
\\
C_{vv}^{(2)}(t)&=-\frac{1}{\pi}\lambda^{-2}-\frac{1}{(2\pi)^{3/2}}\lambda^{-3} +O(\lambda^{-4}).
\end{align}
Therefore, adding the above two results, we recover Jepsen's
result~\cite{jepsen65}
\begin{equation}
C_{vv}(t)=-\frac{(2 \pi-5) }{(2 \pi)^{3/2}}\lambda^{-3}
+O(\lambda^{-4}).
\end{equation}

\subsection{Long time regime}
In the limit $\bar v t >> L$ we integrate Eqs.(\ref{transp}) using the second line of
Eq.(\ref{pxy}): 
\begin{eqnarray}
p_{--}(x, y; t) &=& \frac{x y}{L^2} + f(x, y; t)\nonumber\\
p_{-+}(x, y; t) &=& \frac{x (L - y)}{L^2} - f(x, y; t)\nonumber\\
p_{+-}(x, y; t) &=& \frac{y (L - x)}{L^2} - f(x, y; t)\nonumber\\
p_{++}(x, y; t) &=& \frac{(L - x)(L - y)}{L^2} + f(x, y; t)
\label{jep1-4}
\end{eqnarray}
with
\begin{equation}
f(x, y; t) = \sum_{k\neq 0} \frac{1}{k^2\pi^2}
\exp\left(- \frac{ k^2 \pi^2  \overline v^2 t^2}{2L^2}\right)
\sin\left(\frac{k\pi x}{L}\right) \sin\left(\frac{k\pi y}{L}\right).
\label{jep1-5}
\end{equation}
Expanding around $x = y =L/2$ we get to leading order $N(p_{++}+p_{--})=N[1/2+
2a(t)],~N(p_{++}-p_{--})=-N w_+,~N(p_{+-}+p_{-+})= N[1/2-2a(t)],~N(p_{+-}+p_{-+})=
-N w_-$~,
where $w_+ = (x + y)/L -1,~w_-=(x-y)/L $ and 
\begin{equation}
a(t) = f(L/2, L/2; t) = 
\sum_k \frac{1}{\pi^2 (2 k + 1)^2}
\exp\left[-\frac{(2 k + 1)^2 \pi^2 \overline v^2 t^2 }{2 L^2}\right].
\end{equation}
Using these and the expression of $P_-$ from Eq.~\eqref{P-} in Eqs.~(\ref{v0vt1}) and (\ref{v0vt2}), we find the
following results upto $O(1/N)$:
\begin{align}
C_{vv}^{(1)}(t) = &
N 
 \sum_k \left(1 - \frac{(2k+1)^2\pi^2 \overline v^2 t^2}{L^2}\right)
\exp\left(- \frac{ (2k+1)^2 \pi^2  \overline v^2 t^2}{2L^2}\right)\nn \\
& \qquad\qquad\times
 \f{1}{(2 \pi)^2} ~\int_{-\infty}^\infty {dw_+} \int_{-\infty}^\infty dw_-\int_{-\infty}^{\infty}d \phi \int_{-\infty}^\infty d \theta  
\nn\\
&\qquad\qquad\times
e^{-N [1/4+a(t)] \phi^2}e^{-i N w_+ \phi}  
e^{-N [1/4-a(t)] \theta^2}e^{-i N w_- \theta}  ~,
\label{jep1-2}  \\
C_{vv}^{(2)}(t) = &
-N^2 \left(\f{\bar{v} t}{L}\right)^2 
 \left[ \sum_k \exp\left(- \frac{ (2k+1)^2 \pi^2  \overline v^2 t^2}{2L^2}\right)\right]^2 \nn\\
&\times
 \f{1}{(2 \pi)^2} ~\int_{-\infty}^\infty {dw_+} \int_{-\infty}^\infty 
dw_-  
\int_{-\infty}^{\infty} d \phi \int_{-\infty}^\infty d \theta  
 \nn\\
&\times
e^{-N [1/4+a(t)] \phi^2} e^{-i N w_+ \phi}  
e^{-N [1/4-a(t)] \theta^2}e^{-i N w_- \theta}  ~(\theta^2-\phi^2)~. 
\end{align}
Performing the Gaussian integrals we find that $C_{vv}^{(2)}$
vanishes and hence to $O(1/N)$ the velocity autocorrelation is given
by
\begin{equation}
C_{vv}(t)=C_{vv}^{(1)}(t)
=\frac{2 }{N} 
\sum_{k=1,3,5\ldots}
\left(1 - \frac{k^2\pi^2 \overline v^2 t^2}{L^2}\right)
\exp\left(- \frac{k^2\pi^2\overline v^2 t^2}{2L^2}\right). 
\label{C-long}
\end{equation}
As seen from Fig.~\ref{eqmC1C2}, the above expression describes the
numerical simulation data very well. The late time behaviour $C_{vv}(t) 
\sim  \exp (-\pi^2\overline v^2 t^2/2L^2)$ was earlier obtained in \cite{evans79}.

\section{Simulation results}
\label{sec:simulations}
As mentioned earlier, there are no analytical results when the
particle masses in the one dimensional gas are not all equal. We
turn to numerical simulations for such systems; the simulations
also confirm the analytical results of the previous section, as
shown in Figure~\ref{eqmC1C2}.  The Hamiltonian for the system is
$H = \sum_{l=1}^N  {1\over 2} m_l \dot x_l^2,$ with $0 < x_1 < x_2
\ldots < x_N < L.$ After an elastic collision between two neighboring
particles (say $l$ and $l+1$) with velocities $v_l$, $v_{l+1}$ and
masses $m_l$, $m_{l+1}$ respectively, they emerge with new velocities
$v_l'$ and $v_{l+1}'$. From momentum and energy conservation we
have:
\begin{align}
v_l' &= \frac {(m_l - m_{l+1})} {(m_l + m_{l+1})} v_l + \frac{2 m_{l+1}} {(m_l + m_{l+1})} v_{l+1} \nn \\
v_{l+1}' &= \frac {2 m_l} {(m_l + m_{l+1})} v_l + \frac{(m_{l+1} - m_l)} {(m_l + m_{l+1})} v_{l+1}~.
\end{align}
Between collisions the particles move with constant velocity. 

We simulate this system using an event-driven  algorithm and compute
the  correlation functions $\la [\Delta x(t)]^2 \ra $, $\la \Delta
x(t) v(0) \ra$ and $\la v(t) v(0)$ of the central particle, where
$\Delta x(t)= x_M(t)-x_M(0)$ and $v(t)=v_M(t)$.  The average  $\la
\cdots \ra$ is taken over initial configurations chosen from the
equilibrium distribution, where the  particles are uniformly
distributed in the box with density $\rho=N/L$, while the velocity
of each particle is independently chosen from the distribution $(m/2
\pi k_B T)^{1/2} ~e^{-m v^2/2k_B T}$ .  Note that the three correlation
functions are related to each other as
\begin{displaymath}
\f{1}{2} \frac{d}{dt} \langle [\Delta x(t)]^2\rangle = \la \Delta x(t) v(t) \ra = 
\langle \Delta x(t) v(0)\rangle = \int_0^t \langle v(0)v(t')\rangle dt'= D(t)~.
\end{displaymath}
When the tagged particle shows diffusive behaviour then $\lim_{t
\to \infty} D(t)$ reaches a constant value for an infinite system
and this gives the diffusion constant. On the other hand for sub-diffusion $D(t) $ vanishes as $t \to \infty$  whereas for super-diffusion it diverges. 

Just as for the equal mass system, for any finite system of size
$L$ there is a short time regime during which the tagged particle
at the centre does not feel the effect of the boundaries and during
this time, correlation functions have the same behaviour as the
infinite system.  The time at which the system size effects start
showing up is given by $t_{\rm sat} \sim L/c_s$, where $c_s=\sqrt{3
P/\rho_m}$ is the adiabatic sound velocity  in the hard particle
gas, with  $P$ the pressure and $\rho_m$ the average mass density.
For our numerical simulations, $P = \rho k_BT  = 1$ and $ \rho_m =
1,$ which gives $ c_s = \sqrt 3 $. We now present the results for
the correlation functions in the short-time and long-time regimes.

\begin{figure}
\center
\includegraphics[width=0.7\hsize]{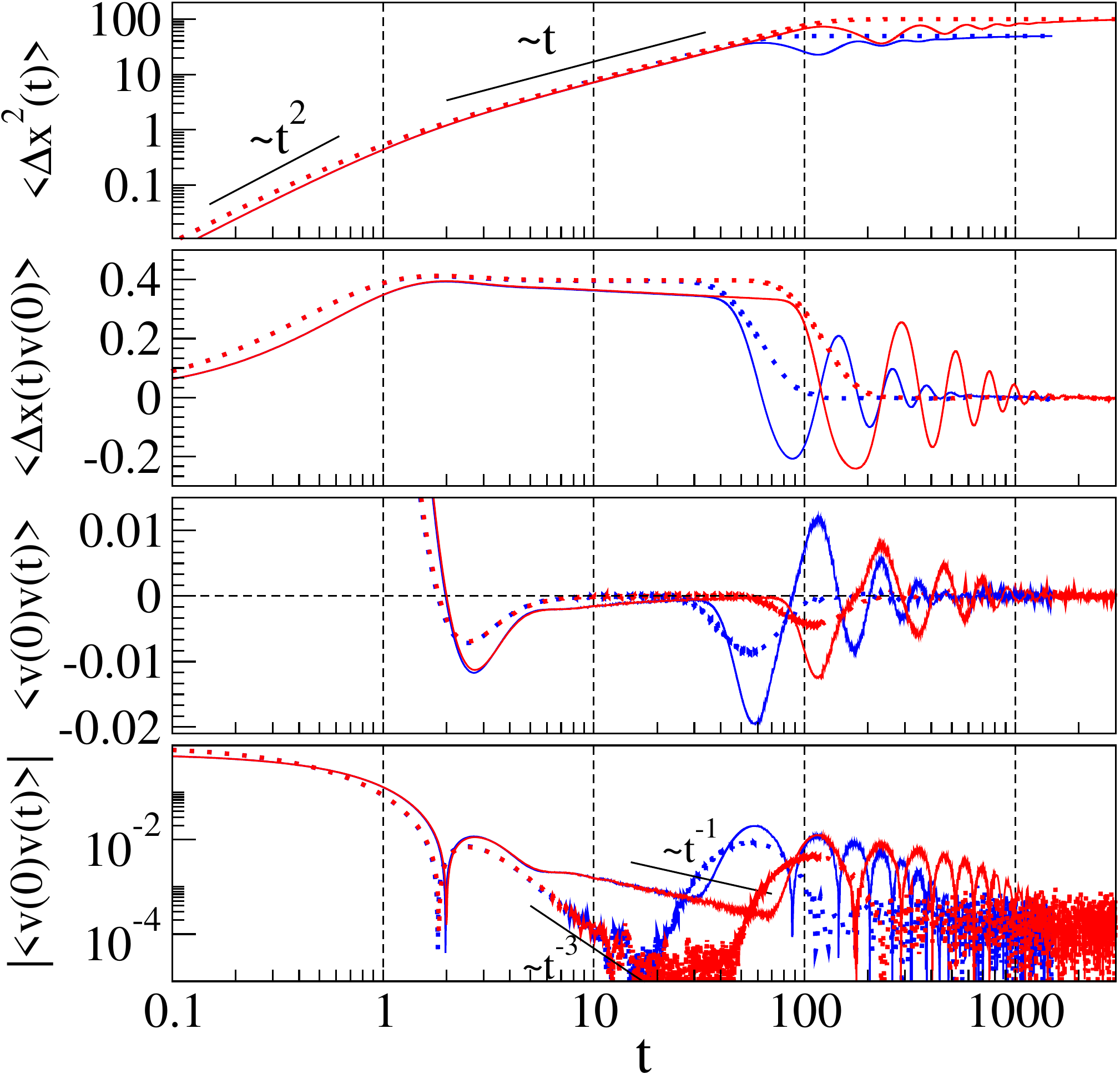}
\caption{\label{altmass} (color online) Various correlation functions for alternate
mass hard particle gas (solid lines) with $N=101$ (blue) and $N=201$ (red) particles, density
$\rho=1$ and $k_B T=1$.  The alternate particles have masses $1.5$ and $0.5$
and in this simulation, the middle particle had mass $1.5$.
The data is obtained by averaging over $10^9$ equilibrium initial
conditions. For comparison, the correlation functions for an equal mass
gas with masses $1$ is also shown (dotted lines).}
\end{figure}

\begin{figure}
\center
\includegraphics[width=0.7\hsize]{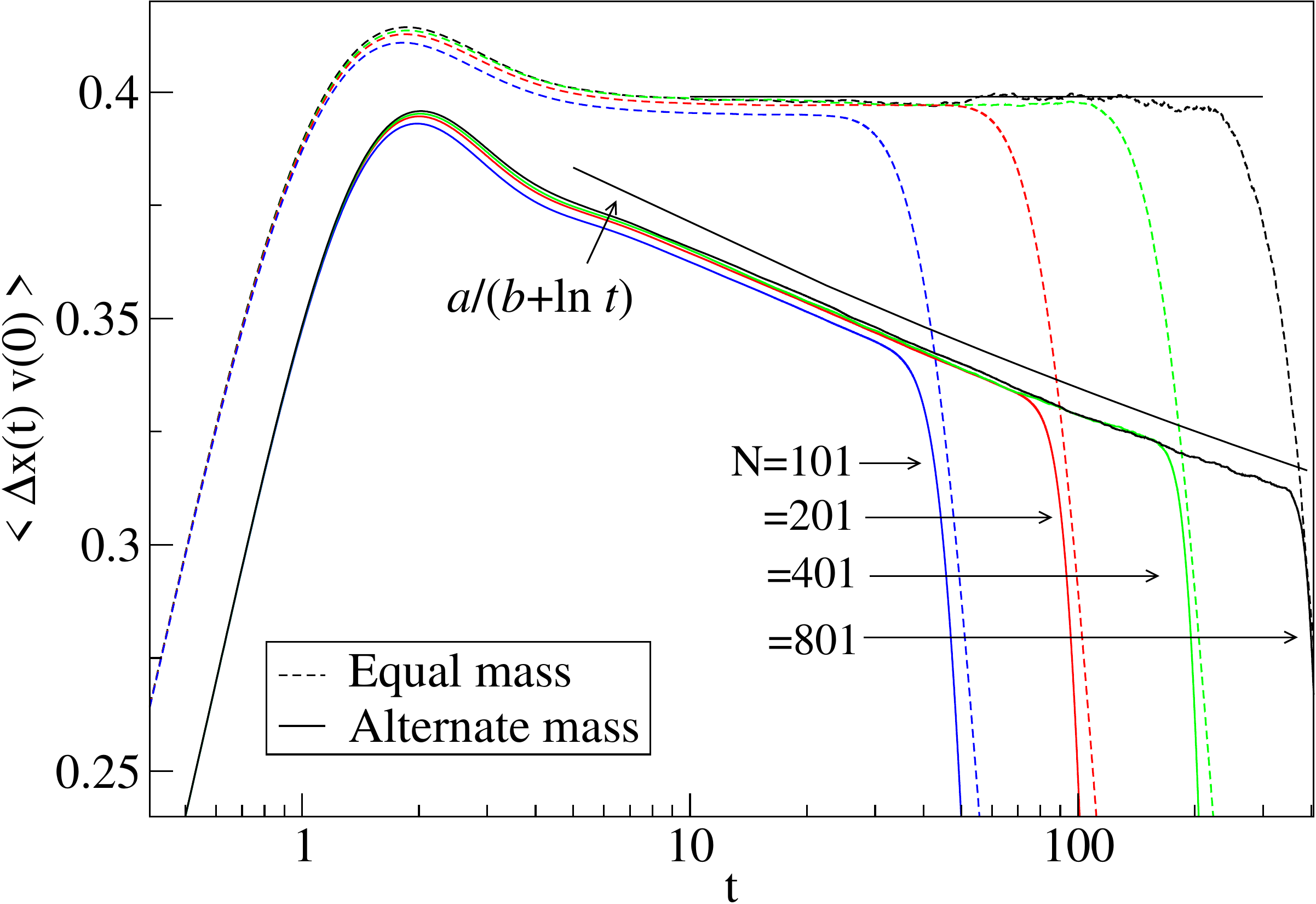}
\caption{\label{xtv0eqalt}  Plot of $D(t)=\la \Delta x(t) v(0) \ra$
  for the alternate mass gas for various system sizes. We see clearly
  the  logarithmic decay of the diffusion constant.  The parameters
  for the shown funciton  are $a=8$ and $b=19$.
  For comparision we also show the corresponding equal mass data
  (dashed line) which shows saturation to the expected Jepsen value
  $1/\sqrt{2 \pi}\approx 0.4$.}
\end{figure}

In Fig.~\ref{altmass}, we show the simulation results for the
correlation functions for a one dimensional hard particle gas with
masses that alternate between $1.5$ and $0.5.$ Here the data is shown
for the case where the tagged particle has mass $1.5$, and similar
results are obtained for the case when the tagged particle is lighter.
For comparison, the results for an equal mass gas are also
shown. After the expected initial ballistic regime, the MSD $\langle
\Delta x^2(t)\rangle$ grows approximately linearly, indicating roughly
diffusive motion.  Simulation results of Marro and
Masoliver~\cite{marro85} obtained $\langle v(0)v(t)\rangle \sim -t^{-
  \delta}$ with $\delta \leq 1$ for the gas with alternating masses,
which would imply (slightly) superdiffusive behavior. It is easiest to
notice any deviations from diffusive behavior in the plot of $\langle
\Delta x(t)v (0)\rangle,$ where diffusive or superdiffusive behavior
would correspond to (after the ballistic regime) a horizontal or
rising straight line respectively. Instead, Figure~\ref{altmass} shows
that $\langle \Delta x(t) v(0)\rangle$ {\it decreases\/} as $t$ is
increased beyond the ballistic regime, implying subdiffusive behavior.
This is seen more clearly in Fig.~(\ref{xtv0eqalt}) where we observe
the dependence $\langle \Delta x(t) v(0)\rangle \sim a/(b+ \ln t)$.
This would correspond to an MSD whose leading part in this regime is
$\sim t/ \ln t$ and a VAF $\sim -1/(t \ln^2 t)$. As seen in
Fig.~(\ref{altmass}) the logarithmic corrections are difficult to
observe in the MSD and VAF.  The smallness of the deviation from
diffusive behavior implies that the apparently linear dependence of
$\langle \Delta x(t) v(0)\rangle$ on $\ln t$, observed in
\cite{marro85}, is nevertheless consistent with the above observed
form for small $\ln t$. However we note that the linear logarithmic
dependence on time as proposed in \cite{marro85} cannot be valid at
large times -- and therefore our form is more appropriate.  Other
functional forms for are also possible.  For instance, $\langle \Delta
x(t) v(0)\rangle \sim t^{-\alpha}$ with a very small $\alpha$ would
also imply the subdiffusive behaviour $\la \Delta x^2(t) \ra \sim
t^{1-\alpha}$.

At long times, the effect of finite size of the box sets in and
the MSD saturates: $\langle [x(t) - x(0)]^2\rangle \rightarrow 2
\langle [x(t) - L/2]^2\rangle,$ which can be easily evaluated  in equilibrium 
to be $L^2/(2N),$ independent of the particle
masses in the gas. We observe this in Fig.~(\ref{altmass}). 
The main difference between  the equal mass and
alternate mass systems is that the MSD for the
equal mass case approaches its saturation value without oscillations,
while for alternate mass case there are damped oscillations as
saturation is approached, \emph{while always remaining below} the MSD for
the equal mass case.  
The oscillations in the alternate mass system
also show up in the other two correlation functions.

\begin{figure}
\center
\includegraphics[width=0.7\hsize]{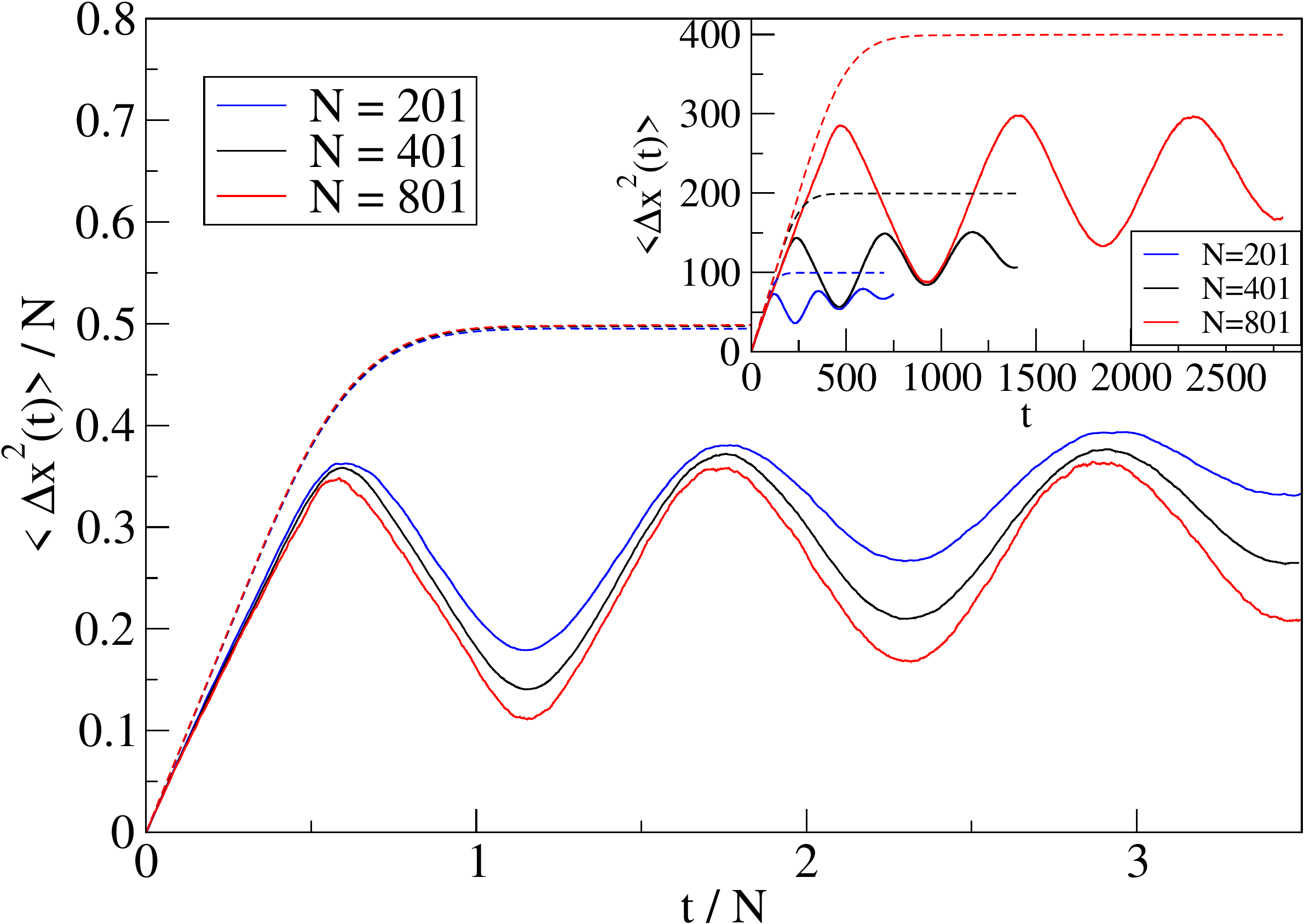}
\caption{ \label{HC2h4h8h} Scaled plot of MSD as a function of time
for three system sizes $N = 201, 401, 801$ for the alternate mass
case with fixed density $\rho=1$ and other parameters as in Fig.~(\ref{altmass}). The inset shows the unscaled data. The saturation value for the scaled plot is at $0.5$.}
\end{figure}

The oscillations in the MSD are seen more clearly in Fig.~\ref{HC2h4h8h},
where the data is plotted differently. The period of oscillation
is proportional to $N,$ in agreement with our discussion earlier
in this section where they were ascribed to sound waves reflecting
from the boundary, which takes a time $\sim L/c_s.$ However, the
amplitude of the oscillations does not show a simple scaling with
$N;$ it is clear from the figure that they are damped out in fewer
cycles for smaller $N,$ making it impossible to collapse the data
onto a single curve by rescaling the vertical axis.

\begin{figure}
\center
\includegraphics[width=0.7\hsize]{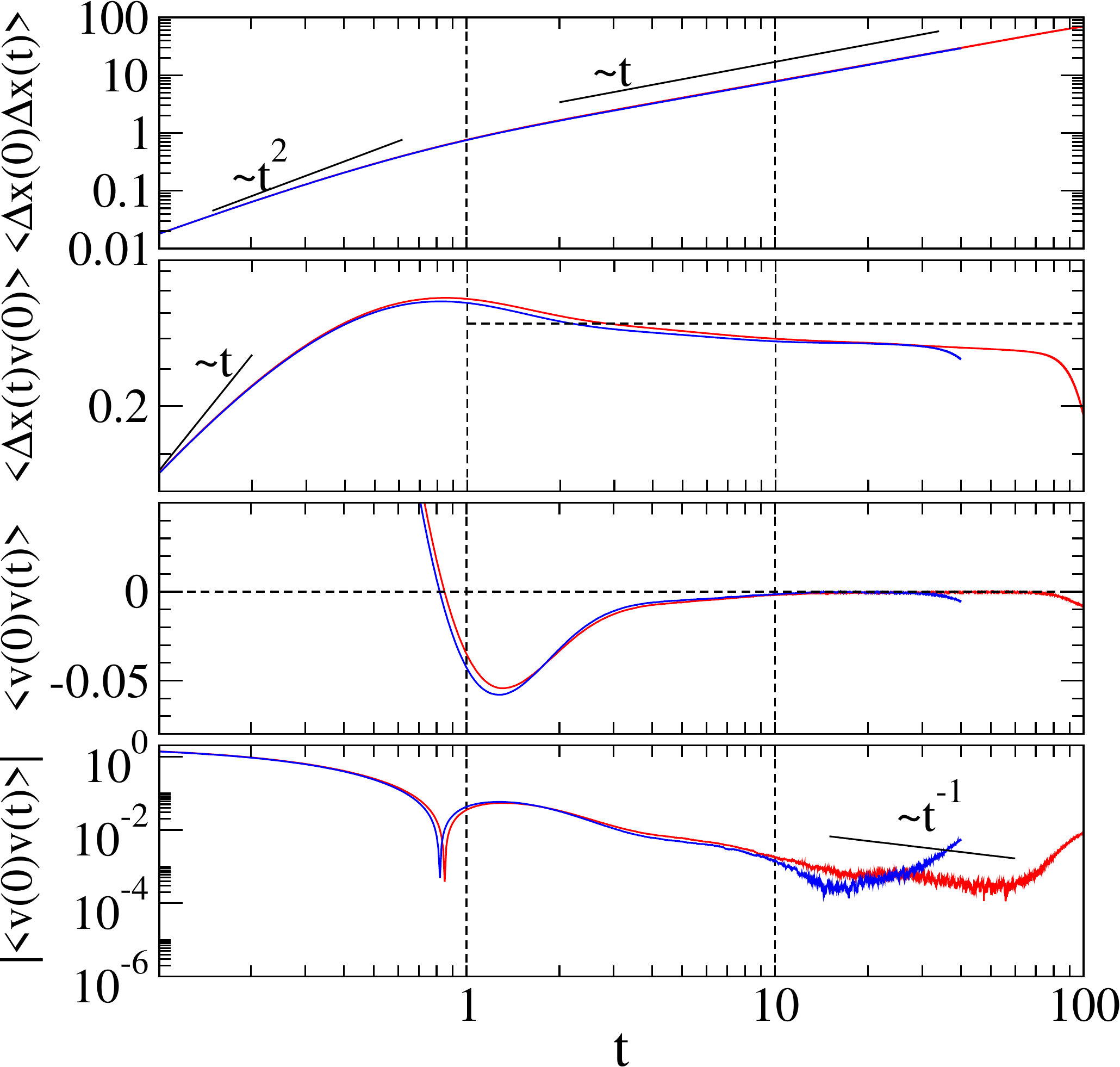}
\caption{\label{randmass} Various correlation functions (in the short-time regime) for random 
mass hard particle gas with $N=101$ and $N=201$ particles and density
$\rho=1$. The mass of the middle particle was always taken to be 0.5 and the results are an average over $32$ different random mass realizations.} 
\end{figure}

To check for the robustness of our results we have also performed simulations 
of  a gas with random distribution of masses. Each particle was assigned a 
mass  from a uniform distribution between  $0.5$ and $1.5$. We looked at tagged-particle correlations of the central particle whose mass was fixed at $0.5$. 
The correlations fluctuate between different mass realizations and we took an average over $32$ realizations. The results are plotted in Fig.~(\ref{randmass})
where we see the same qualitative features as for the alternate mass case.

\section{Summary}
\label{sec:summary}
We have studied tagged particle correlations of the middle particle in a system of $N$ hard point particles confined in a one-dimensional box of length $L$ and in thermal equilibrium. For the case where the masses of all particles are equal we obtained 
analytic results for the finite-size velocity auto-correlation function using the approach of Jepsen. 
We have presented a somewhat simpler and physically motivated  calculation of the velocity auto-correlation function
 and 
obtained closed form expressions valid at both short times (including the ballistic and diffusive regimes) and long times (when finite size effects show up). 
While here we have only presented results for the velocity auto-correlations, 
it  is straightforward to obtain other correlation functions using our approach.

Next we have presented simulation results for the case of a hard-point gas where the particles have unequal masses. Two cases are studied, one where particles 
have alternate masses and the other where the masses are random. In both cases 
we find that the behaviour of correlation functions is qualitatively different from the equal mass case. 
The correlation $\la \Delta x(t) v(0) \ra$ does not saturate to a 
constant (expected for a diffusive behaviour) and  instead shows a slow decay consistent with the form $\la \Delta x^2(t)\ra  \sim t/\ln t$.
Correspondingly the VAF  decays  as $\sim 1/(t \ln^2 t) $ which is
completely different from the equal mass form $\sim 1/t^3$.  This indicates that
tagged-particle motion is sub-diffusive.  
However it is difficult to see this sub-diffusive behaviour directly in the 
mean square displacement  of the tagged particle since the deviation from linear time-dependence is small.
These results are surprising since simulations with other interacting systems such as Lennard-Jones gases have found diffusive motion and $1/t^3$ decay of the velocity auto-correlation function \cite{bishop81}. Understanding this difference as well as studying tagged particle motion in 
 other interacting systems and higher dimensional systems remain interesting open problems.

\appendix
\section{Details of calculation}
\label{appendix}
In this appendix, we provide a more detailed calculation of the velocity autocorrelation function for the hard particle gas of equal mass. This is an alternative to the derivation of some of the key equations in this paper.
To compute $\langle v_M(0) v_M(t)\rangle_1$, we pick at time $t=0$ one
of the non-interacting particles at random from the distribution
$p(x_0,v_0)$. At time $t$ let the position and velocity of the
particle be given by $x_t(x_0,v_0)$ and $v_t(x_0,v_0)$
respectively. We then calculate the probability, $P_N^{(1)}(x_0,x_t)$,
that it has an equal number of particles to its left and right at both
the initial and final times, \emph{i.e.}, at $t=0$ and $t$.  For
$\langle v_M(0) v_M(t)\rangle_2$, we pick two non-interacting
particles from the distribution $p(x_0,v_0)
~p(\tilde{x}_0,\tilde{v}_0)$ and let them evolve to $x_t,v_t$ and
$\tilde{x}_t,\tilde{v}_t$ respectively. We then calculate the
probability $P_M^{(2)}(x_0,\tilde{x}_t)$ that at time $t=0$, the first
particle $x_0$ is the middle particle while at time $t$ the second
particle $\tilde{x}(t)$ is the middle particle. The normalized VAF is
thus given by:
\begin{align}
C_{vv}^{(1)}(t) & =\f{\langle v_M(0) v_M(t)\rangle_1}{\bar{v}^2} = \f{N}{\bar{v}^2}  \int_{0}^{L} \frac{dx_0}{L} \int_{-\infty}^\infty 
\frac{dv_0}{\sqrt{2\pi}\,\overline v}
e^{- v_0^2/2 \overline v^2}\, v_0 v_t\, P_{N}^{(1)} (x_0,x_t,t)~, \label{c1-1} \\
C_{vv}^{(2)}(t)&=\f{\langle v_M(t) v_M(0)\rangle_2}{\bar{v}^2} \nn \\ 
&= \f{N (N - 1)}{\bar{v}^2} \idotsint \frac{dx_0}{L} \frac{d\tilde x_0}{L} 
\frac{dv_0 d\tilde v_0}{2\pi\overline v^2}\, v_0 \tilde v_t \,
e^{-(v_0^2+\tilde v_0^2)/2\overline v^2}~
P_{N}^{(2)} (x_0,x_t,\tilde x_0,\tilde x_t,t)~. \label{c2-1}
\end{align}
These forms together with the explicit expressions of $P_N^{(1),(2)}$ discussed below, agree with those given in \cite{lebowitz72}.
We now make a change of variables from $x_0,v_0$ to 
$x_0,x_t$ in Eq.~(\ref{c1-1}) and from $x_0,v_0,\tilde{x}_0,\tilde{v}_0$ to 
$x_0,x_t,\tilde{x}_0,\tilde{x}_t$ in Eq.~(\ref{c2-1}). 

In the non-interacting picture, $x_t$ and $v_t$, as well as the number
of collisions $m$, suffered by the particle with the walls upto time
$t$, are completely determined by the initial configuration
$(x_0,v_0)$.  The number of collisions with the wall is given by
\begin{equation}
m=\begin{cases}\displaystyle
\left\lfloor \f{x_0+v_0 t}{L} \right\rfloor &\quad \text{if}~v_0 > 0,\\[3mm]
\displaystyle
\left\lfloor \f{L-x_0-v_0 t}{L} \right\rfloor &\quad \text{if}~v_0 < 0~,
\end{cases}
\end{equation}
where $\lfloor x \rfloor$ is the integral part of $x$.
When $m$ is even, we have $v_t=v_0$ whereas $v_t=-v_0$ for odd $m$.
The final position $x_t$ is given by one of the following relations
depending on $m$ and $v_0$. When $m$ is even, we have
$x_0+v_0t=mL+x_t$ for $v_0>0$ and $L-x_0-v_0 t= mL +L-x_t$ for
$v_0<0$.  On the other hand for odd $m$ we get $x_0+v_0t=mL+L-x_t$ for
$v_0>0$ and $L-x_0-v_0 t=mL + x_t$ for $v_0<0$.  Combining all these
four cases, we can write $x_0 + v_0 t = 2 n L \pm x_t$. Here $n=m/2$
and $-m/2$ respectively for the first two cases where $m$ is even and
the plus sign is taken. For the last two cases, where $m$ is odd,
$n=(m+1)/2$ and $-(m+1)/2$ respectively and the minus sign is taken.
In other words, for a given values of $x_0$ and $v_0$ in the relations
$x_0 + v_0 t = 2 n L \pm x_t$ and $v_t=\pm v_0$, the values of $n$ and
$x_t$, and the signs taken from the $\pm$ are uniquely
determined. Therefore, inserting the term $[\delta(x_0+v_0
  t-2nL-x_t)\delta(v_t-v_0) +\delta(x_0+v_0
  t-2nL+x_t)\delta(v_t+v_0)]$ in the integrand of Eq.~\eqref{c1-1}
while integrating over $x_t$ and $v_t$, and summing over all integer
values of $n$, does not change the result, i.e., 
\begin{multline}
\langle v_M(t) v_M(0)\rangle_1 = 
N \int_0^L dx_t  \int_{-\infty}^\infty dv_t \sum_{n=-\infty}^\infty 
\int_0^L \frac{dx_0}{L} \int_{-\infty}^\infty 
\frac{dv_0}{\sqrt{2\pi}\,\overline v}
 e^{- v_0^2/2 \overline v^2}\,v_0 v_t\, P_{N}^{(1)}(x_0,x_t,t)  \\
\times \bigl[\delta(x_0+v_0
  t-2nL-x_t)\delta(v_t-v_0) +\delta(x_0+v_0
  t-2nL+x_t)\delta(v_t+v_0)\bigr].
\label{vv-1}
\end{multline}
Now, carrying out the integrations over $v_t$ and $v_0$, after some
straightforward manipulation we obtain
\begin{align}
C_{vv}^{(1)} = 
N \int_{0}^{L} {dx_0} \int_{0}^{L} {dx_t}
~P_{N}^{(1)}(x_0,x_t,t)
\,\overline v ~\partial_{\overline v}~ P_-(x_0,x_t,t),
\label{jep1-1a} ~.
\end{align}

The second part of the velocity autocorrelation function is given by
Eq.~\eqref{c2-1} and in this case we trade the $v_0, \tilde v_0$
integrals for $x_t,\tilde x_t$ by introducing two sets of
$\delta$-function, one for each particle as in Eq~\eqref{vv-1}. After
some manipulations we then get 
\begin{align}
C_{vv}^{(2)} &= N(N-1)
\idotsint {dx_0} {d\tilde x_0} 
{dx_t d\tilde x_t} ~P_{N}^{(2)}(x_0,\tilde{x}_0,x_t,\tilde{x}_t) \nn \\
&~~~~~~~~~~~~~~~~~~~~~\bigl[\overline v t~\partial_{x_0} P(x_0,x_t)\bigr] 
\bigl[-\overline v t~ \partial_{\tilde x_t}  P(\tilde x_0,\tilde
  x_t)\bigr], \label{jep2-1}
\end{align}

{\bf Evaluation of $P_N^{(1)}(x_0,x_t)$}:
This gives the probability that, at $t=0$ and at time $t$, the selected 
particle has an equal number of particles to its left and right.
We note that the remaining $N-1$
particles are independent of each other and the selected particle. Let
$p_{-+}(x_0, x_t; t)$ be the probability that one of these particles
is to the left of $x_0$ at $t=0$ and to the right of $x_t$ at time
$t$. Let $p_{+-}$, $p_{--}$ and $p_{++}$ be similarly defined. In
terms of these probabilities, it is easily seen that
\begin{equation}
P_{N}^{(1)} =
\sum_{n_1+n_2+ n_3+ n_4=N-1} \frac{(N-1)!}{n_1! n_2! n_3! n_4!} \,
p_{--}^{n_1} p_{-+}^{n_2} p_{+-}^{n_3} p_{++}^{n_4}\,
\delta_{n_1, n_4} \delta_{n_2, n_3},
\label{P_(N-1)/2^(1)}
\end{equation}
where in the summand, $n_1$ particles go from the left of $x_0$ to the
left of $x_t$, $n_2$ particles from the left to the right, $n_3$
particles from the right to the left, and $n_4$ particles from the
right to the right.  The two Kronecker delta functions ensure that an
equal number of particles cross the selected particle in both
directions in time $t$ and that an equal number of particles remain on
either side of the selected particle.  Together, these conditions are
equivalent to an equal number of particles being on either side of the
selected particle at time $0$ and $t$, that is, $n_1+n_2 = n_3 + n_4$
and $n_1+n_3=n_2+n_4$.  The multinomial coefficient takes care of all
possible permutations among the particles.  Now, using the integral
representation of the Kronecker delta,
\begin{math}
\delta_{m,n}=(2\pi)^{-1} \int_0^{2\pi} e^{i(m-n)\theta} \, d\theta 
\end{math}
in the above equation immediately gives Eq.~\eqref{jep1-3}

{\bf Evaluation of $P_N^{(2)}(x_0,x_t)$}: In calculating $P_{N}^{(2)}$
we have to keep track of both the particles. There arise four
situations: (a) $x_0 > \tilde x_0$ and $x_t < \tilde x_t$, (b) $x_0 >
\tilde x_0$ and $x_t > \tilde x_t$, (c) $x_0 < \tilde x_0$ and $x_t >
\tilde x_t$, and (d) $x_0 < \tilde x_0$ and $x_t < \tilde x_t$. Let
there be $n_1$ particles go from the left of $x_0$ to the left of
$\tilde x_t$, $n_2$ particles from the left to the right, $n_3$
particles from the right to the left, and $n_4$ particles from the
right to the right.  Since two of the particles are considered
separately, the rest can be chosen $(N-2)!/(n_1! n_2! n_3!  n_4!)$
different ways and $n_1+n_2+n_3+n_4=N-2$. Now, in the first situation
we have (a) $n_1+n_2+1=n_3+n_4$ and $n_1+n_3+1=n_2+n_4$. These
conditions are equivalent to $n_2=n_4$ and $n_1=n_4-1$. Similarly one
can work out the conditions for the other three situations which gives
(b) $n_1=n_4$ and $n_2=n_3-1$, (c) $n_2=n_3$ and $n_1=n_4+1$, and (d)
$n_1=n_4$ and $n_2=n_3+1$, respectively.  Following the procedure used
to evaluate $P_{N}^{(1)}$, we can easily find $P_{N}^{(2)}$ as given
by Eq.~\eqref{jep2-2}, 
where the extra phase factor $\psi(\theta,\phi)$ originates from
addend $\pm1$ that appear in the relations among $n_i$'s above, and
$\psi(\theta,\phi)=e^{-i\phi}$, $e^{-i\theta}$, $e^{i\phi}$ and
$e^{i\theta}$ for situations (a), (b), (c) and (d) respectively.

{\bf Evaluation of $P(x_0,x_t)$}:
The joint probability density function for a
(non-interacting) particle to be between $x$ and $x+dx$ at $t=0$ and
between $y$ and $y+dy$ at time $t$ is given by
\begin{align}
P(x,y) &= \la \delta (x-x_0) \delta(y-x_t) \ra \nn \\
 &=\int_0^L \f{dx_0}{L} \int_{-\infty}^\infty dv_0~  \f{e^{-v_0^2/2\bar v^2}}{\sqrt{2 \pi} \bar v}~ \delta (x-x_0) \delta(y-x_t) \nn \\
&= \int_0^L dx_t  \int_{-\infty}^\infty dv_t \sum_{n=-\infty}^\infty 
\int_0^L \f{dx_0}{L} \int_{-\infty}^\infty dv_0  ~\f{e^{-v_0^2/2\bar v^2}}{\sqrt{2 \pi} \bar v}~ \delta (x-x_0) \delta(y-x_t)  ~ \nn \\
& \times 
  \bigl[\delta(x_0+v_0
  t-2nL-x_t)~\delta(v_t-v_0) +\delta(x_0+v_0
  t-2nL+x_t)~\delta(v_t+v_0)\bigr].
\end{align}
Now, carrying out the integrations over all the variables gives the
first line of Eq.~\eqref{pxy}.

\end{document}